\documentclass[aps,twocolumn,prb,superscriptaddress,amsmath,amssymb]{revtex4-1}
\usepackage{graphicx}
\usepackage{dcolumn}
\usepackage{bm}
\usepackage{float}
\usepackage{placeins}
\usepackage{hyperref}
\usepackage{hhline}
\usepackage{multirow}
\usepackage{array}
\usepackage{tabularx}
\usepackage{booktabs}
\usepackage{amsmath}
\usepackage{ifthen}
\usepackage[dvipsnames]{xcolor}
\usepackage{soul}
\hypersetup{colorlinks=true,linkcolor=blue,filecolor=magenta,urlcolor=blue,citecolor=blue,pdftitle={Overleaf Example}}

\newcommand{\point}[1]%
{%
    \ifthenelse{\equal{#1}{1}}{$E$}{}%
    \ifthenelse{\equal{#1}{2}}{$3^{1}_{(001)}$}{}%
    \ifthenelse{\equal{#1}{3}}{$3^{2}_{(001)}$}{}%
    \ifthenelse{\equal{#1}{4}}{$2_{(110)}$}{}%
    \ifthenelse{\equal{#1}{5}}{$2_{(100)}$}{}%
    \ifthenelse{\equal{#1}{6}}{$2_{(010)}$}{}%
    \ifthenelse{\equal{#1}{7}}{$\mathcal{I}$}{}%
    \ifthenelse{\equal{#1}{8}}{$\bar{3}^{1}_{(001)}$}{}%
    \ifthenelse{\equal{#1}{9}}{$\bar{3}^{2}_{(001)}$}{}%
    \ifthenelse{\equal{#1}{10}}{$\mathcal{M}_{(110)}$}{}%
    \ifthenelse{\equal{#1}{11}}{$\mathcal{M}_{(100)}$}{}%
    \ifthenelse{\equal{#1}{12}}{$\mathcal{M}_{(010)}$}{}%
}
\newcommand{\trans}[1]%
{%
    \ifthenelse{\equal{#1}{1}}{$\mathcal{T}_{000}$}{}%
    \ifthenelse{\equal{#1}{2}}{$\mathcal{T}_{100}$}{}%
    \ifthenelse{\equal{#1}{3}}{$\mathcal{T}_{010}$}{}%
    \ifthenelse{\equal{#1}{4}}{$\mathcal{T}_{110}$}{}%
    \ifthenelse{\equal{#1}{5}}{$\mathcal{T}_{001}$}{}%
    \ifthenelse{\equal{#1}{6}}{$\mathcal{T}_{101}$}{}%
    \ifthenelse{\equal{#1}{7}}{$\mathcal{T}_{011}$}{}%
    \ifthenelse{\equal{#1}{8}}{$\mathcal{T}_{111}$}{}%
}
\newcommand{\symm}[2]{$\{$\point{#1}$\,|\,$\trans{#2}$\}$}

\begin{document}

\title{Origin of chirality in transition-metal dichalcogenides}
\author{Kwangrae~Kim}
\altaffiliation{These authors contributed equally to this work.}
\affiliation{Department of Physics, Pohang University of Science and Technology, Pohang 37673, South Korea}
\author{Hyun-Woo J.~Kim}
\altaffiliation{These authors contributed equally to this work.}
\affiliation{Department of Physics, Pohang University of Science and Technology, Pohang 37673, South Korea}
\affiliation{Advanced Photon Source, Argonne National Laboratory, Lemont, IL 60439, USA}
\author{Seunghyeok~Ha}
\altaffiliation{These authors contributed equally to this work.}
\affiliation{Department of Physics, Pohang University of Science and Technology, Pohang 37673, South Korea}
\author{Hoon~Kim}
\affiliation{Department of Physics, Pohang University of Science and Technology, Pohang 37673, South Korea}
\author{Jin-Kwang~Kim}
\affiliation{Department of Physics, Pohang University of Science and Technology, Pohang 37673, South Korea}
\author{Jaehwon~Kim}
\affiliation{Department of Physics, Pohang University of Science and Technology, Pohang 37673, South Korea}
\author{Hyunsung~Kim}
\affiliation{Department of Physics, Pohang University of Science and Technology, Pohang 37673, South Korea}
\author{Junyoung~Kwon}
\affiliation{Department of Physics, Pohang University of Science and Technology, Pohang 37673, South Korea}
\author{Jihoon~Seol}
\affiliation{Department of Physics, Pohang University of Science and Technology, Pohang 37673, South Korea}
\author{Saegyeol~Jung}
\affiliation{Center for Correlated Electron Systems, Institute for Basic Science, Seoul 08826, South Korea}
\affiliation{Department of Physics and Astronomy, Seoul National University, Seoul 08826, South Korea}
\author{Changyoung~Kim}
\affiliation{Center for Correlated Electron Systems, Institute for Basic Science, Seoul 08826, South Korea}
\affiliation{Department of Physics and Astronomy, Seoul National University, Seoul 08826, South Korea}
\author{Ahmet Alatas}
\affiliation{Advanced Photon Source, Argonne National Laboratory, Lemont, IL 60439, USA}
\author{Ayman Said}
\affiliation{Advanced Photon Source, Argonne National Laboratory, Lemont, IL 60439, USA}
\author{Michael Merz}
\affiliation{Institute for Quantum Materials and Technologies, Karlsruhe Institute of Technology, Karlsruhe 76021,Germany}
\affiliation{Karlsruhe Nano Micro Facility, Karlsruhe Institute of Technology, Eggenstein-Leopoldshafen 76344,Germany}
\author{Matthieu Le Tacon}
\affiliation{Institute for Quantum Materials and Technologies, Karlsruhe Institute of Technology, Karlsruhe 76021,Germany}
\author{Jin Mo~Bok}
\email[email: ]{jinmobok@postech.ac.kr}
\affiliation{Department of Physics, Pohang University of Science and Technology, Pohang 37673, South Korea}
\author{Ki-Seok~Kim}
\email[email: ]{tkfkd@postech.ac.kr}
\affiliation{Department of Physics, Pohang University of Science and Technology, Pohang 37673, South Korea}
\author{B.~J. Kim}
\email[email: ]{bjkim6@postech.ac.kr}
\affiliation{Department of Physics, Pohang University of Science and Technology, Pohang 37673, South Korea}

\date{\today}

\maketitle

\noindent
{\bf Chirality is a ubiquitous phenomenon in which a symmetry between left- and right-handed objects is broken, examples in nature ranging from subatomic particles and molecules to living organisms. In particle physics, the weak force is responsible for the symmetry breaking and parity violation in beta decay, but in condensed matter systems interactions that lead to chirality remain poorly understood. Here, we unravel the mechanism of chiral charge density wave formation in the transition-metal dichalcogenide 1$T$-TiSe$_2$. Using representation analysis, we show that charge density modulations and ionic displacements, which transform as a continuous scalar field and a vector field on a discrete lattice, respectively, follow different irreducible representations of the space group, despite the fact that they propagate with the same wave-vectors and are strongly coupled to each other. This charge-lattice symmetry frustration is resolved by further breaking of all symmetries not common to both sectors through induced lattice distortions, thus leading to chirality. Our theory is verified using Raman spectroscopy and inelastic x-ray scattering, which reveal that all but translation symmetries are broken at a level not resolved by state-of-the-art diffraction techniques.
}

In quantum solids, chirality is defined by the absence of inversion centers, roto-inversion axes and mirror/glide planes in their space group symmetries. Chirality endows electrons with diverse exotic physical properties, such as circular photogalvanic effect (CPGE) \cite{asn79}, giant anomalous Hall effect \cite{tag01}, non-reciprocal transport \cite{tok18} and chiral anomaly \cite{son13,xio15}, with potential applications to spintronic devices and energy harvesting technology. However, the mechanism whereby chirality emerges from the constituent microscopic degrees of freedom remains unknown, and our inability to predict chiral electronic phases from first principles severely limits fully exploiting their potentially useful properties. 

\begin{figure*}
\centering
\includegraphics[width=1.85\columnwidth]{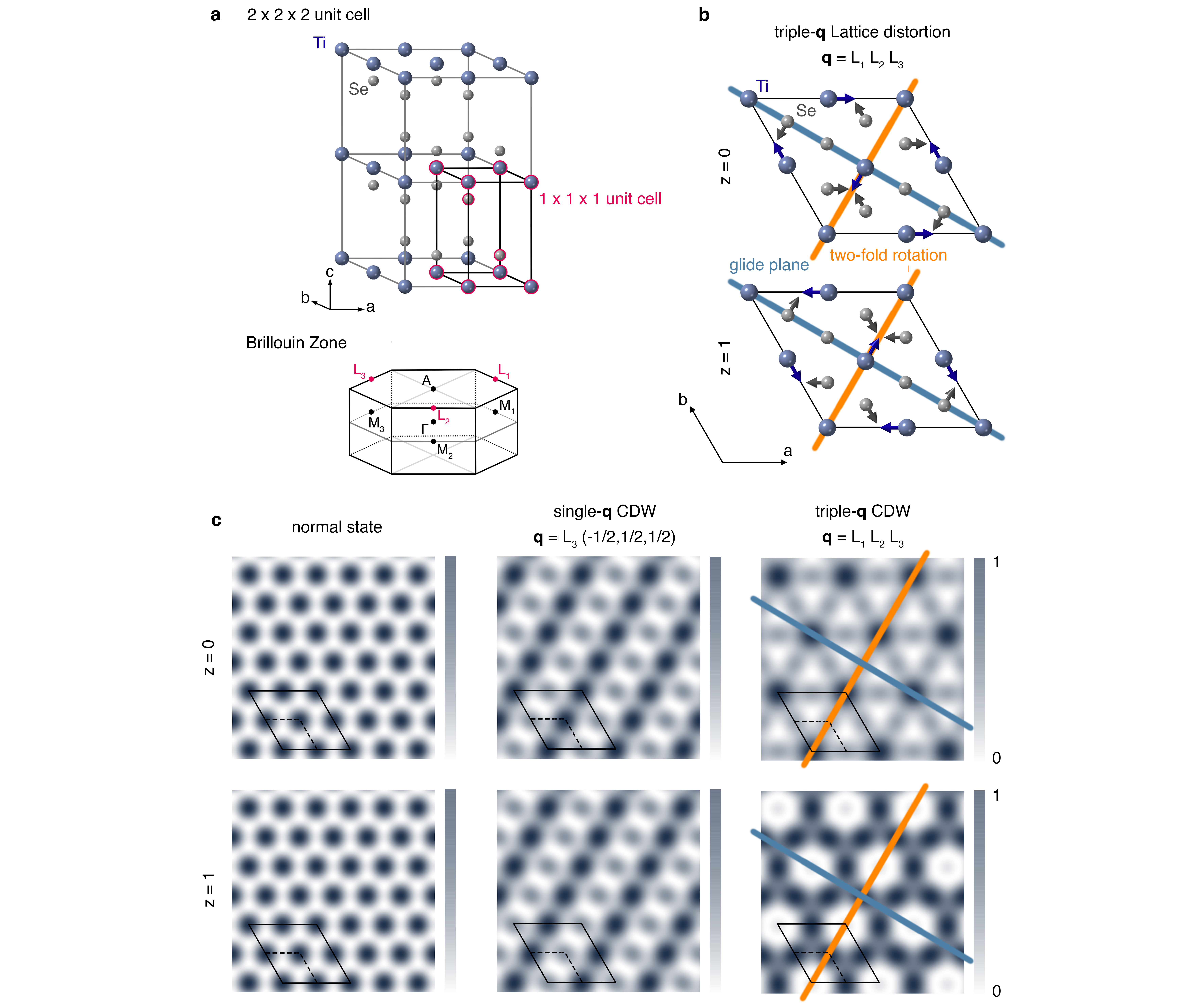}
\caption{\textbf{Charge density wave and 2\,$\times$\,2\,$\times$\,2 superstructure.} $\mathbf{a}$, 2\,$\times$\,2\,$\times$\,2 enlarged unit cell and Brillouin zone of 1$T$-TiSe$_2$. $\mathbf{b}$, Lattice distortions in the triple-$\mathbf{q}$ CDW phase. $\mathbf{c}$, Charge density for the normal phase, single-$\mathbf{q}$ CDW phase, and triple-$\mathbf{q}$ CDW phase. Orange and blue lines are two-fold rotation axis and glide plane, respectively. There are three two-fold rotation axes and glide planes related by three-fold rotation, but only one of each are shown for clarity.}
\label{fig1}
\end{figure*}

The notion that chirality can arise from a multi-component CDW, an electronic order of scalar quantity, has opened a new avenue to explore chiral properties of quantum matters. Based on a phenomenological theory, it has been shown that CDW acquires chirality, defined by $\mathbf{q}_1\cdot\mathbf{q}_2\times\mathbf{q}_3$, upon dephasing of waves propagating along three different wave-vectors ($\mathbf{q}_1$, $\mathbf{q}_2$, $\mathbf{q}_3$) \cite{ish10,wez11}. Experimental evidence for chiral CDWs is primarily based on scanning tunneling spectroscopic data on transition-metal dichalcogenides of chemical formula AB$_2$ (A=Ti, Ta, V and B=S, Se) hosting triple-$\mathbf{q}$ CDWs, which exhibit different charge modulation intensities in tunneling currents measured along three equivalent crystallographic directions \cite{ish10,ish11,pas19,yan22,sin22}. Similar observations have been made in a different family of kagome materials AV$_3$Sb$_5$ (A=K, Rb, Cs), which point to a general mechanism for chirality in triple-$\mathbf{q}$ CDW materials \cite{jia21,shu21,wan21}. 

A major argument against chirality, however, is based on the fact that no evidence for a lattice symmetry breaking consistent with the putative chiral CDW phase has ever been found \cite{fan17,lin19,ued21,zha22}. More puzzling is the recent observation of CPGE in 1$T$-TiSe$_2$, a macroscopic manifestation of chirality, which is forbidden in its known crystal structure \cite{xu20}; Neumann's principle states that the symmetry of any macroscopic property must include all symmetry elements of the lattice \cite{ore21}. Is the lack of structure evidence merely due to our inability to resolve subtle lattice distortions even with modern synchrotron-based x-ray diffraction? Or, does it signify possible violation of the fundamental principle and a profound flaw in our understanding of structure-property relationship in quantum matters?

Although widely taken for granted, it is not trivial that ions on a discrete lattice are always able to precisely follow the symmetry of a continuous charge distribution of valence electrons, given the different transformation properties innate to scalar (charge) and vector (lattice) quantities. In this Article, we perform a rigorous symmetry analysis employing the full space group of 1$T$-TiSe$_2$. % an archetypal CDW material with a long history of research over four decades. 
We show that the symmetry inconsistency is resolved by successive reduction of the symmetries of the two sectors until they become equivalent. In this process, all of the mirror, inversion, and roto-inversion symmetries present in the system are lost. %, thus providing a route to chirality.
As a result, the system becomes chiral at the onset of the CDW, although neither the charge density modulation nor the atomic displacement is chiral by itself. We confirm this result by using inelastic x-ray scattering and Raman spectroscopy. We discuss implications of our result  on the nature of chiral symmetry breaking phase transitions and possible violation of the Neumann's principle.

\noindent
\textbf{Charge-lattice symmetry frustration}

Below the transition temperature $T_C$\,$\approx200$ K, the triple-$\mathbf{q}$ CDW sets in with a 2\,$\times$\,2\,$\times$\,2 modulation of the charge density \cite{sal76,woo76} (Fig.~1). The ordering wave-vectors are $\mathbf{q}_{L_1}$\,=\,$(1/2,\,0,\,1/2)$, $\mathbf{q}_{L_2}$\,=\,$(0,\,-1/2,\,1/2)$ and $\mathbf{q}_{L_3}$\,=\,$(-1/2,\,1/2,\,1/2)$, which connect the hole Fermi pocket centered at the zone center ($\Gamma$) and the electron pockets at the three $L$ points related by three-fold rotation symmetry \cite{ros01,wat19}. Concomitantly, lattice distortions result from condensation of $ab$-plane polarized phonon modes of the same $\mathbf{q}$'s \cite{sal76,bia15,hol01,web11}. 

The order parameters in the charge ($\delta \rho$) and the lattice ($\mathbf{d}$) sectors are written as

\begin{align}
    \delta\rho&=\delta\rho_{L_1}+\delta\rho_{L_2}+\delta\rho_{L_3}, \\
    \mathbf{d}&=\mathbf{d}_{L_1}^T+\mathbf{d}_{L_2}^T+\mathbf{d}_{L_3}^T,
\end{align}

\noindent where

\begin{align}
    \delta\rho_{L_i}&=\Re\, e^{i\mathbf{q}_{L_i}\cdot\mathbf{r} + \varphi_i}, \\
    \mathbf{d}_{L_i}^T&=\Re\, {\boldsymbol\epsilon}_{i} e^{i\mathbf{q}_{L_i}\cdot \mathbf{r} + \varphi_i}.
\end{align}

Note that $\mathbf{d}$ has the same functional form as $\delta \rho$ apart from the polarization vector ${\boldsymbol\epsilon}_i$, which is transverse to $\mathbf{q}_i$ (${\boldsymbol\epsilon}_i\cdot\mathbf{q}_i=0$) \cite{wez11}. $\varphi_i$'s give the initial phases for each component at the origin. The CDW acquires chirality when the three components propagate out of phase with one another, i.e.  $\varphi_1$\,$\neq$\,$\varphi_2$\,$\neq$\,$\varphi_3$ \cite{wez11}. However, the chiral symmetry in the lattice sector cannot be broken with the displacements as given by Eq.~(4): the lowest possible space group symmetry is $P\bar{1}$ (Table 1), which includes inversion and thus is always achiral regardless of $\varphi_i$ values. %This difference stems from the scalar(vector) nature of $\delta \rho$ ($\mathbf{d}_{L}$). 

In fact, the space groups are different even for the achiral CDW phase with $\varphi_1$\,=\,$\varphi_2$\,=\,$\varphi_3$. Upon inspection of the patterns of $\delta \rho$ and $\mathbf{d}_{L}^T$ shown in Fig.~1, it is clear that the mirror planes are replaced by glides by $\mathbf{d}_{L}^T$, changing the space group from $P\bar{3}m1$ to $P\bar{3}c1$, corresponding to the known high and low temperature structures, whereas they are preserved by $\delta \rho_L$. These considerations challenge the widespread notion that charge and lattice sectors should have the same symmetry.

\noindent
\textbf{Symmetry analysis} 

For a systematic analysis, we solve for the irreducible representations (IRs) of the space group $P\bar{3}m1$ and classify possible $\delta \rho$ and $\mathbf{d}$ in terms of them. Because both $\delta \rho$ and $\mathbf{d}$ have the periodicity of the 2\,$\times$\,2\,$\times$\,2 unit cell, it is sufficient to consider the factor group $\mathcal{G}/\mathcal{T}$, where $\mathcal{G}$ is the space group $P\bar{3}m1$ and $\mathcal{T}$ is the infinite translation group generated by $\mathcal{T}_{200}$, $\mathcal{T}_{020}$, and $\mathcal{T}_{002}$, which translate the system by twice the unit lattice vectors. $\mathcal{G}/\mathcal{T}$ is a finite group of order 96, and has eight one-dimensional (1D) IRs, four 2D IRs, and eight 3D IRs. We provide the full analysis in the Supplementary Information and summarize here only the key results. Here, we focus on four relevant 3D IRs, $\Gamma_1^-$, $\Gamma_4^+$, $\Gamma_5^-$, $\Gamma_7^-$.

\begin{figure*}
\centering
\includegraphics[width=1.85\columnwidth]{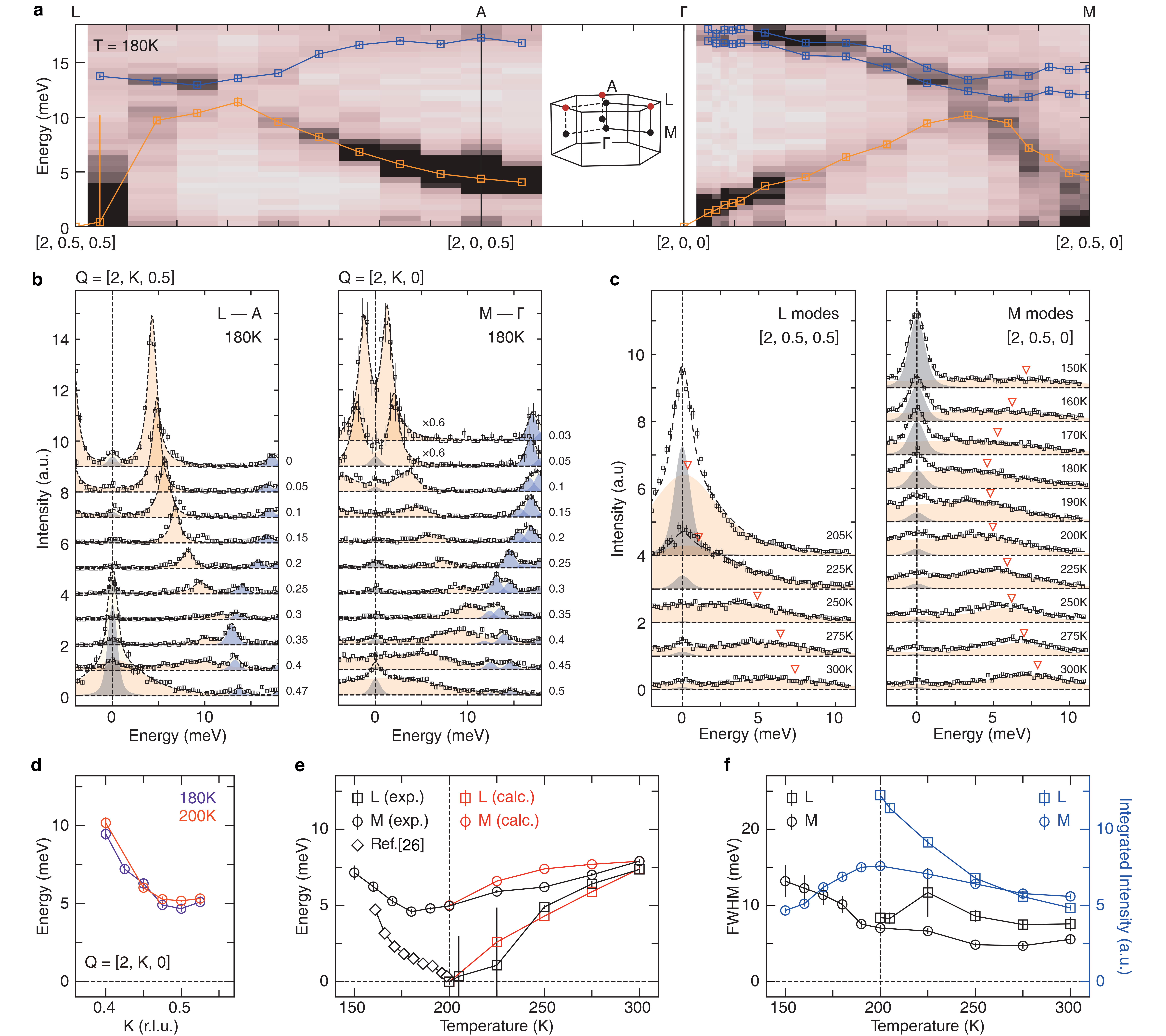}
\caption{\textbf{Inelastic x-ray scattering.} $\mathbf{a}$, Phonon dispersion along $L-A-\Gamma-M$ measured at $T$=180 K. Orange and blue squares represent the peak centers of acoustic and optical modes, respectively. $\mathbf{b}$, Spectra along $L-A$ and $M-\Gamma$ at 180 K. Orange and blue shaded regions are for acoustic and optical modes, respectively. $\mathbf{c}$, Temperature dependent spectra at $L$ and $M$. Orange triangles are the peak centers of the acoustic mode. $\mathbf{d}$, Dispersions of the acoustic phonon at 180 K and 200 K near $M$. $\mathbf{e}$, Temperature dependence of mode energies at $L$ and $M$. Diamond markers are phonon energies extracted from thermal diffuse scattering \cite{hol01}. $\mathbf{f}$, Black markers represent the temperature dependence of full-width-half-maximum (FWHM). Blue markers represent integrated intensity of energy above 1 meV. All spectra are fitted with Lorentzian function with Bose factor and error bars indicate one standard deviation.}
\label{fig2} 
\end{figure*}

\begin{figure*}
\centering
\includegraphics[width=1.85\columnwidth]{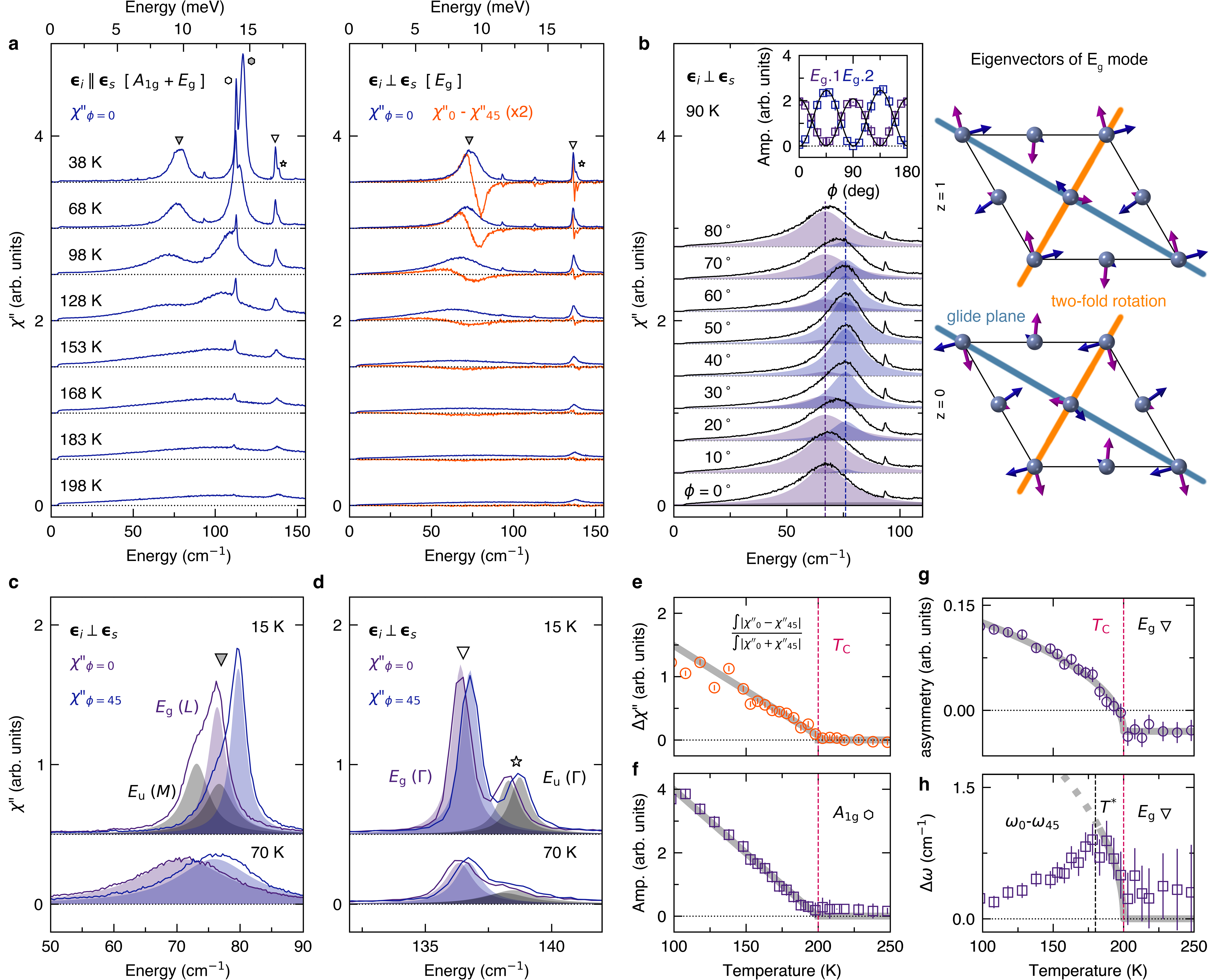}
\caption{\textbf{Raman scattering.} $\mathbf{a}$, Temperature dependence of Raman spectra measured with parallel (left panel) and perpendicular (right panel) incident and scattered light polarizations. Orange lines are differences between the spectra obtained at $\phi$=0$^\circ$ and at $\phi$=45$^\circ$. $\mathbf{b}$, Polarization angle ($\phi$) dependence of Raman spectra at $T$ = 90 K. Dashed lines indicate the peak centers determined from $\phi$=0$^\circ$ and $\phi$=45$^\circ$ spectra. Inset displays the peak amplitudes, which are fitted (black lines) by the Raman tensor for $C_{2h}$ point group. Eigenvectors for $E_\mathrm{g}$ modes are shown when they are degenerate. $\mathbf{c,d}$, $E_\mathrm{g}$ and $E_\mathrm{u}$ modes at 15 K and 70 K. $\mathbf{e-h}$, Temperature dependence of ($\mathbf{e}$) the measure of splitting ($\mathbf{f}$) zone-folded $A_\mathrm{1g}$ mode amplitude ($\mathbf{g}$) asymmetry of $E_\mathrm{g}$ mode at 135 cm$^{-1}$ ($\mathbf{h}$) splitting of $E_\mathrm{g}$ mode at 135 cm$^{-1}$. Error bars indicate one standard deviation.}
\label{fig3}
\end{figure*}
 
Table 1 summarizes the key difference in the symmetry transformation properties between $\delta \rho$ and $\mathbf{d}$. When $\varphi_1$\,=\,$\varphi_2$\,=\,$\varphi_3$\,=\,0, they belong to $\Gamma_4^+$ and $\Gamma_7^-$ IRs, respectively, the former (latter) of which is even (odd) under the space inversion about the origin. From their different IRs, it is clear that $\delta\rho$ and $\mathbf{d}$ break different sets of symmetry elements. Then, the symmetry of the system as a whole can only be as high as that of a group containing only the common symmetry elements preserved by both $\delta \rho$ and $\mathbf{d}$. The isotropy subgroups of $\Gamma_4^+(a,a,a)$ and $\Gamma_7^-(a,a,a)$, triple-$\mathbf{q}$ states with same amplitudes and phases (equal to zero), share only six symmetry elements, in which inversion, glide, and roto-inversion symmetries are absent. Thus, although neither $\delta \rho$ or $\mathbf{d}$ is chiral by itself, the system as a whole becomes chiral.    

Accordingly, the lattice symmetry has to be lowered. In fact, the simultaneous presence of $\delta\rho$ and $\mathbf{d}$ necessarily induces lattice distortions that transform as the combined symmetry of $\delta\rho\,\mathbf{d}$. In particular, $\Gamma_4^+ \otimes\Gamma_7^-$ contains $\Gamma_5^-$, which represents the transverse phonon modes at the $M$ point ($\mathbf{d}_{M}^T$) (Supplementary Tables 18 and 28). Thus, allowed in the Landau free energy are terms that couple linearly to $\mathbf{d}_{M}^T$, of the form $\delta\rho_{L_1}\mathbf{d}_{L_2}^T\mathbf{d}_{M_3}^T$ plus its permutation of the indices, which always lower the free energy with nonzero $\mathbf{d}_{M}^T$ regardless of the sign of the coupling constant (Supplementary Table 3 and Supplementary Fig.~1). With a nonzero $\mathbf{d}_{M}^T$, the highest symmetry space group is $P321$, which is further lowered to $P1$ when other induced displacements of 1D and 2D IRs are included (Supplementary Table 2). Indeed, our experimental data presented below are only consistent with $P2$ or $P1$.

\noindent
\textbf{Inelastic x-ray scattering}

A verifiable prediction of our analysis is that there must be only one structural phase transition at $T_\mathrm{C}$\,$\approx 200$ K, where chiral symmetry becomes broken. In particular, we predict that there is no structural transition associated with condensation of the $M$ modes, despite being heavily renormalized by CDW fluctuations, as the transition is preempted by the induced $\mathbf{d}_{M}^T$. This is in contrast to the earlier predictions based on Landau theory which predicts that the chiral CDW phase is reached on cooling through two phase transitions \cite{wez11,cas13}: (i) $\varphi_{1}$\,=\,$\varphi_{2}$\,=\,$\varphi_{3}$ for $T^*$\,$<$\,$T$\,$<$\,$T_\mathrm{C}$, and (ii) $\varphi_{1}$\,=\,0, $\varphi_{2}$\,=\,$-\varphi_{3}$\,$\neq$\,0 for $T$\,$<$\,$T^*$, for which anomalies in transport and specific heat data suggest $T^*$$\sim$ 180 K \cite{cas13}.

Figures 2a and 2b show the inelastic x-ray scattering spectra along the high symmetry lines $L-A-\Gamma-M$, measured at $T$\,=\,180 K near $T^*$, in a transmission geometry sensitive to $ab$-plane transverse $L$ and $M$ modes. We measure only one of the three modes from $L_1$, $L_2$, and $L_3$ (and $M_1$, $M_2$, and $M_3$), because three-fold rotation symmetry guarantees that their spectra are identical above $T_\mathrm{C}$. 
Along $A-L$, a well-defined peak at $\sim$4 meV at $A$ disperses up in energy and bends down as $L$ is approached. The phonons are heavily damped near $L$, indicative of strong coupling to CDW fluctuations. A similar trend is found along $\Gamma-M$; the lineshapes become extremely broad as $M$ is approached. At this temperature, $L$ modes have already condensed, as is well established from earlier studies \cite{hol01,web11}, but $M$ modes remain at finite energies. Figures 2c-2f show the temperature evolution of $L$ and $M$ modes. At both $L$ and $M$, elastic peaks appear below $T_\mathrm{C}$, which are Bragg reflections associated with the formation of the supercell. Whereas $L$ modes soften upon cooling, reaching zero energy at $T_\mathrm{C}$ from $\sim$7 meV at room temperature, $M$ modes remain at finite energies and harden below $T^*$. This can also be seen from the large increase of the integrated intensity (due to the Bose factor) observed for $L$ modes when they condense, but not for $M$ modes (Fig.~2f). 

The similar behaviors of $L$ and $M$ modes are due to the quasi-2D nature of the electronic band structure, which expects similar amplitudes for $\Gamma \leftrightarrow L$ and $\Gamma \leftrightarrow M$ charge fluctuations, renormalizing the phonon dispersions near $L$ and $M$, respectively. Based on DFT electronic band structures and random phase approximation (RPA), we reproduce the softening of $L$ and $M$ modes (Fig.~2a and 2e) using a three-parameter model (electron-hole exchange coupling, electron-phonon couplings for $L$ and $M$ modes) (details in the Supplementary Information). Our model takes into account the charge harmonics belonging to different IRs, without which fitting to experimental data requires much larger coupling constants (Supplementary Figs.~4-8). Our central claim, however, only depends on the fact that there is no condensation of $M$ modes at or below $T_\mathrm{C}$, where our Raman data presented below show clear evidence for symmetry breaking below $P\bar{3}c1$. This can only be consistent with induced freezing of $M$ modes, because a $2\times 2\times 2$ reconstruction can only involve $L$, $A$, $\Gamma$, and/or $M$ modes, and there are no anomalies at $A$ and $\Gamma$ (Supplementary Fig.~9).

\noindent
\textbf{Raman scattering}

Because the Bragg peaks at $L$ and $M$ grow rapidly with decreasing temperature and dwarf inelastic signals, it is difficult to follow the phonon modes to low temperatures using IXS. Instead, they become visible in Raman spectra as they are back-folded into the zone center. 
Using polarization-resolved Raman spectroscopy, we present direct evidence of chiral symmetry breaking at $T_\mathrm{C}$. 
The point group of the space group is $D_{3d}$ for both $P\bar{3}m1$ and $P\bar{3}c1$ space groups, which allows two species of Raman-active modes belonging to $A_\mathrm{1g}$ and $E_\mathrm{g}$ IRs. Therefore, the three-fold degenerate $L$ and $M$ modes split into a singlet and a doublet as they are back-folded. Their symmetries can be identified from the fact that $A_\mathrm{1g}$ ($E_\mathrm{g}$) modes are allowed only when the incident and scattered photon polarizations are parallel to each other (visible in both parallel and cross polarization configurations) (Supplementary Fig.~2) \cite{lou01}.

First, we show that $E_\mathrm{g}$ doublets further split into singlets, which implies breaking of three-fold rotation symmetry. Figure 3a shows the temperature dependence of Raman spectra. Slightly below $T_\mathrm{C}$, the $E_\mathrm{g}$ doublet originating from back-folding of $L$ modes are seen as a broad hump spanning energy range between 0$\sim$10 meV as in the IXS spectra. Upon cooling, it becomes a well-defined peak, though still much broader than other phonon modes, and hardens to $\sim$75 cm$^{-1}$, consistent with the $L$ mode identified in earlier Raman studies \cite{sug80,sno03}. 

Figure 3b shows its polarization angle ($\phi$\,=\,0$^\circ$ when the incident photon is polarized along the $a$-axis) dependence measured at $T$\,=\,90 K. The peak center shifts as $\phi$ is rotated, which means that the doublet is split into two singlets ($E_\mathrm{g}.1$ and $E_\mathrm{g}.2$) that trade intensities with each other. Their intensities oscillate sinusoidally as a function of $\phi$ out of phase with each other, which is a behavior expected for an $E_\mathrm{g}$ doublet. As they transform into each other by glide and two-fold rotations (Fig.~3b), their splitting implies absence of these symmetries and that the symmetry of the lattice is lower than $P\bar{3}c1$. The highest space group symmetry consistent with this observation is $C2/c$; a necessary and sufficient condition for $E_\mathrm{g}$ mode splitting is breaking of three-fold rotation symmetry, which implies breaking of at least one of the three glide planes (and also two-fold rotations) related by the same symmetry (details in the Supplementary Information). 

Next, we determine the temperature at which the $E_\mathrm{g}$ mode splitting occurs. To this end, we plot in Fig.~3a the difference of the spectra measured at $\phi$\,=\,0$^\circ$ and $\phi$\,=\,45$^\circ$, where the intensity of the one of the two modes becomes maximal and the other minimal. The difference spectrum clearly shows the splitting as a peak and a dip, and thus integrating its absolute value provides a measure for the symmetry breaking. From its temperature dependence plotted in Fig.~3e, it is clear that the onset of the splitting matches with $T_\mathrm{C}$, which is independently measured by the intense $A_\mathrm{1g}$ mode appearing at $\sim$120 cm$^{-1}$ with the zone folding (Fig.~3f). This result shows that the crystal structure with the space group $P\bar{3}c1$, known as the low-temperature phase for more than four decades, is not realized at any temperature.     

Upon a close inspection of the base temperature spectra shown in Fig.~3c, we resolve also $M$ mode nearly degenerate with $L$ mode. At $\phi$\,=\,0$^\circ$ and $\phi$\,=\,45$^\circ$, where one of the split doublet has vanishing intensity, we observe that the spectra still have asymmetric lineshapes, which can be well fitted with two peaks of similar widths having the same $\phi$ dependence. The $M$ mode, however, has odd parity (Table 1), and thus its presence in the Raman spectra implies breaking of the inversion symmetry. This is further confirmed by the presence of infrared-active $E_\mathrm{u}$ mode at $\sim$138 cm$^{-1}$, as shown in Fig.~3d \cite{sub22,hol77}. As the peak broadens at higher temperatures (Supplementary Fig.~3), it merges with a nearby mode $E_\mathrm{g}$ at $\sim$135 cm$^{-1}$, a zone-center mode in the original unit cell visible at all temperatures. Thus, fitting the spectra with a single peak with an asymmetric lineshape, we find that the asymmetry goes to zero at $T_\mathrm{C}$ (Fig.~3g), indicating inversion symmetry breaking at the onset of the CDW order. Thus, removing inversion symmetry from $C2/c$, space groups consistent with our Raman data are limited to $P1$, $P2$, $C2$ and $Cc$. Combined with the IXS result that $\mathbf{d}_{M}^T$ is induced at $T_\mathrm{C}$, possible low-temperature phase can only be either $P2$ or $P1$. $P2$ is further ruled out if we consider other phonon modes (not measured in this study) that must also be frozen (Supplementary Table 2). 

\noindent
\textbf{Discussion}

Our work comprehensively unravels the mechanism of chiral symmetry breaking in 1$T$-TiSe$_2$. Theoretically, provided that charge modulations and lattice distortions are given by Eq.~(1) and Eq.~(2), respectively, it follows deductively that their simultaneous presence necessarily breaks chiral symmetry due to their different symmetry transformation properties. Ever since the pioneering work by McMillan \cite{mcm75}, Eq.~(1) is assumed in all theoretical studies that we are aware of. But is it a priori obvious that the charge modulations can only be expressed as Eq.~(1)? For example, if the phase transition is primarily driven by a lattice instability and charge modulations merely follow it, their spatial distributions would have the form of $\sin\mathbf{G}_{L}\cdot\mathbf{r}\cos\mathbf{q}_{L}\cdot\mathbf{r}$ (Table 1). In other words, the lowest-order plane-wave components in the Fourier expansion of the charge modulations would be $\exp{i (\mathbf{G}_{L}+\mathbf{q}_{L})\cdot\mathbf{r} }$ (Supplementary Fig.~11). However, this possibility is ruled out by the presence of superlattice peaks in the first Brillouin zone in the STM data \cite{ish10,hil18}. Then, the only remaining possibility not considered so far is the charge modulations of the form $\sin\mathbf{q}_{L}\cdot\mathbf{r}$ belonging to $\Gamma_1$ IR (Table 1). In this case, a cascade of symmetry breaking results following a similar logic but through different phonon modes. However, the lowest possible symmetry would then be $P\bar{1}$, which is ruled out by our Raman data. 

%Without reference to any theory, our experimental data alone clearly reveal chiral symmetry breaking at the onset of CDW. 
With the rigorous group theoretical analysis the space group is uniquely determined to be $P1$. This is a highly unusual case where all but translation symmetries are broken at a second order transition. A possible scenario is a phase transition driven by a chiral electronic instability, but this is unlikely considering the small density of states at the Fermi level.  
Instead, our work explains how the usual CDW instability, driven cooperatively by electron-lattice coupling \cite{wez10,zen13}, results in a chiral symmetry breaking without chiral instability in either the charge or the lattice sector. Further, our RPA calculation starting from an effective field theory respecting $P\bar{3}m1$ space group symmetry reproduces the experimental data with reasonable parameters, which otherwise require much larger values to offset the weak Fermi surface nesting (Supplementary Figs.~4-8). More importantly, our formulation shows that the polarization bubble with generic momentum dependence can dynamically couple charge and lattice modes despite their different symmetries, which explains how two symmetry-distinct order parameters can arise at a single phase transition. This point is discussed in detail in the Supplementary Information. 

%Because our mechanism relies only on the generally different symmetry transformation properties of ionic displacements and charge modulations, it may also account for chirality in other multi-$\mathbf{q}$ CDW materials \cite{jia21,shu21,wan21}. 

A remaining question is: why does CPGE appear only below $T^*$? As the chirality scales with $\mathbf{d}_{M}^T$, which in turn grows in proportion to $\delta\rho$, the feedback between the CDW order and the chiral lattice distortions amplifies each other as the temperature is lowered. Thus, although the chiral lattice symmetry is broken at $T_\mathrm{C}$, chirality grows slowly with decreasing temperature, and it may take some further cooling before it becomes detectable. In other scenario, a competing phase nearly degenerate with the ground state may be responsible for CPGE. We note that the $E_\mathrm{g}$ mode splitting decreases below $T^*$ (Fig.~3h), which can be interpreted as a tendency to restore three-fold rotation symmetry. In fact, a recent study based on DFT and optical pump-probe experiment suggests that a $P321$ structure may be stabilized in a non-equilibrium condition \cite{wic22,har23}. 

Finally, we comment on the possible violation of the Neumann's principle. Coming back to the earlier question: is the lattice always able to follow the symmetry of a charge modulation? In our case, the charge-lattice symmetry frustration is resolved through induced lattice distortions in the lowest-order coupling of $\delta\rho_L$, $\mathbf{d}_{L}^T$, and $\mathbf{d}_{M}^T$. However, it is not clear that phonon modes of an appropriate symmetry will generally be available in other systems. 
Formally, the frustration will eventually be resolved through higher-order couplings, but in our case the induced lattice distortions in the lowest order is already subtle enough to evade detection for many decades. 
Thus, in every practical sense, it may be possible to find a ``purely'' electronic transition decoupled from the lattice if the symmetry mismatch is not resolved in the lowest order.

\begin{table*}
\centering
%\begin{ruledtabular}
\begin{tabular}{|c|c|c|l|l|}
\hline
\multirow{2}{*}{\textbf{IR}} & \multicolumn{2}{c|}{\textbf{basis}} & \multirow{2}{*}{\textbf{ subgroups}} & \multirow{2}{*}{\textbf{ symmetry elements}} \\ 
\cline{2-3}
& \textbf{charge} & \textbf{lattice} & & \\
%Gamma1
\hline
\multirow{14}{*}{$\;\Gamma_1^-\;$} & \multirow{2}{*}{$\sin\mathbf{q}_{L_3}\cdot\mathbf{r}$} & \multirow{2}{*}{${\boldsymbol\epsilon}_1\cos\mathbf{q}_{L_3}\cdot\mathbf{r}$} & \multirow{2}{*}{$\;(aaa)$: $P\bar{3}m1\;$} & \multirow{2}{*}{$\;$\symm{1}{1}; \symm{2}{1}; \symm{3}{1};}\\
& & & & \\
& \multirow{6}{*}{\raisebox{-0.5\totalheight}{\includegraphics[width=0.25\columnwidth]{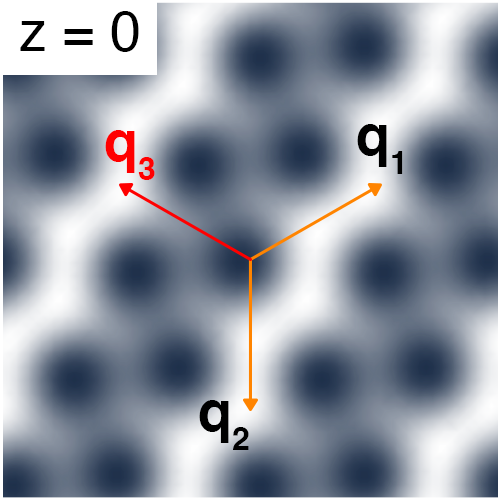}}} & \multirow{6}{*}{\raisebox{-0.5\totalheight}{\includegraphics[width=0.25\columnwidth]{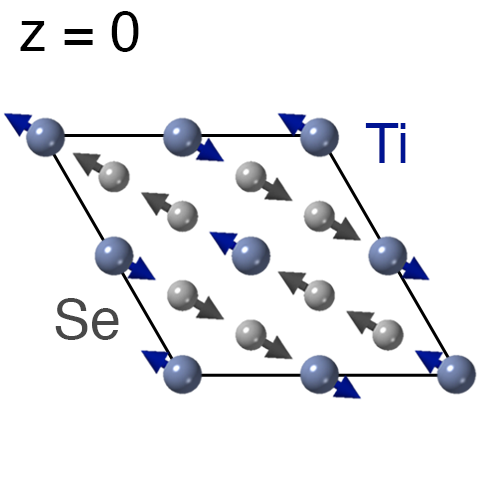}}} & & \multirow{2}{*}{$\;$\symm{10}{1}; \symm{11}{1}; \symm{12}{1};}\\
& & & & \\
& & & & \multirow{2}{*}{$\;$\symm{4}{5}; \symm{5}{5}; \symm{6}{5};}\\
& & & & \\
& & & & \multirow{2}{*}{$\;$\symm{7}{5}; \symm{8}{5}; \symm{9}{5}}\\
& & & & \\
& \multirow{6}{*}{\raisebox{-0.5\totalheight}{\includegraphics[width=0.25\columnwidth]{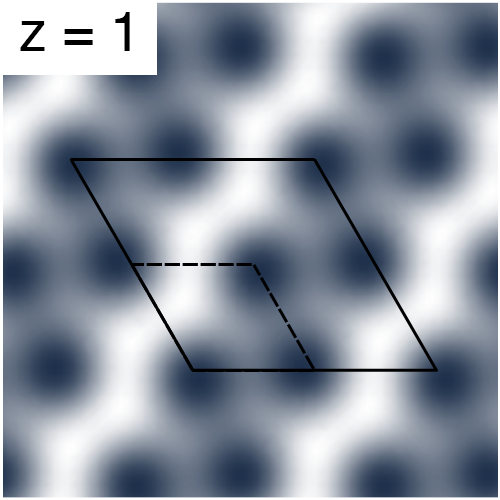}}} & \multirow{6}{*}{\raisebox{-0.5\totalheight}{\includegraphics[width=0.25\columnwidth]{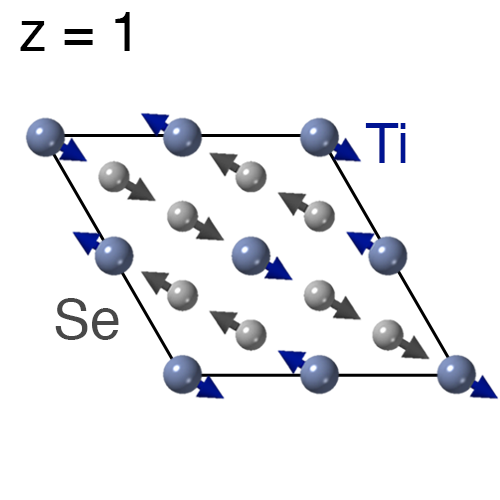}}} & \multirow{2}{*}{$\;(aab)$: $C2/m\;$} & \multirow{2}{*}{$\;$\symm{1}{1}; \symm{11}{1}; \symm{5}{5}; \symm{7}{5}}\\
& & & & \\
& & & \multirow{2}{*}{$\;(abc)$: $P\bar{1}\;$} & \multirow{2}{*}{$\;$\symm{1}{1}; \symm{7}{5}} \\
& & & & \\
& & & & \\
& & & & \\
%Gamma4
\hline
& \multirow{2}{*}{$\cos\mathbf{q}_{L_3}\cdot\mathbf{r}$} & \multirow{2}{*}{${\boldsymbol\epsilon}_4\cos\mathbf{q}_{L_3}\cdot\mathbf{r}$} & \multirow{2}{*}{$\;(aaa)$: $P\bar{3}m1\;$} & \multirow{2}{*}{$\;$\symm{1}{1}; \symm{2}{1}; \symm{3}{1};}\\
& & & & \\
& \multirow{6}{*}{\raisebox{-0.5\totalheight}{\includegraphics[width=0.25\columnwidth]{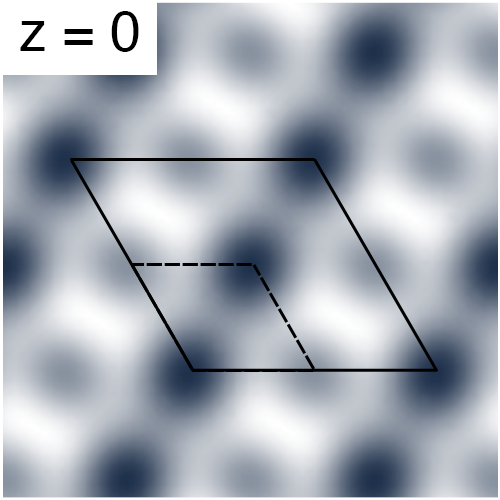}}} & \multirow{6}{*}{\raisebox{-0.5\totalheight}{\includegraphics[width=0.25\columnwidth]{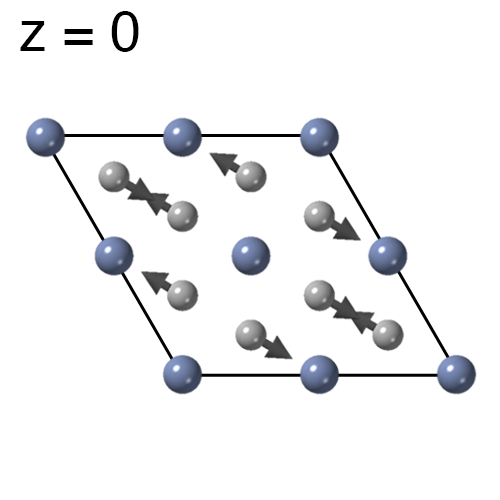}}} & & \multirow{2}{*}{$\;$\symm{4}{1}; \symm{5}{1}; \symm{6}{1};}\\
& & & & \\
& & & & \multirow{2}{*}{$\;$\symm{7}{1}; \symm{8}{1}; \symm{9}{1};}\\
& & & & \\
\multirow{1}{*}{$\;\Gamma_4^+\;$} & & & & \multirow{2}{*}{$\;$\symm{10}{1}; \symm{11}{1}; \symm{12}{1}}\\
\multirow{2}{*}{$(\delta\rho_L)$} & & & & \\
& \multirow{6}{*}{\raisebox{-0.5\totalheight}{\includegraphics[width=0.25\columnwidth]{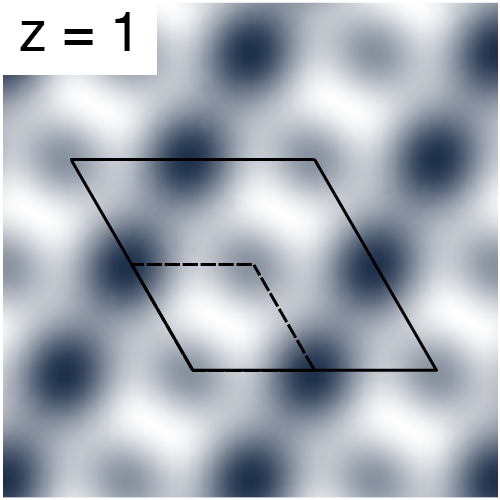}}} & \multirow{6}{*}{\raisebox{-0.5\totalheight}{\includegraphics[width=0.25\columnwidth]{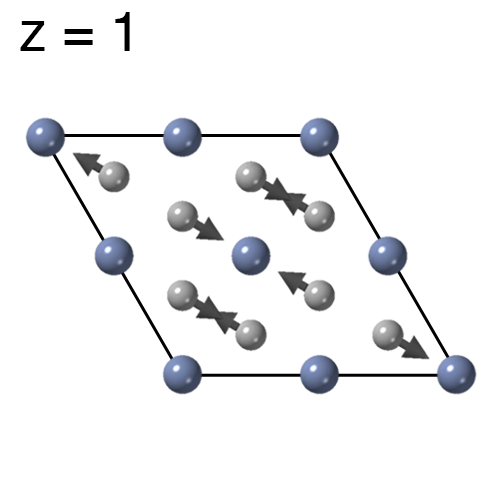}}} & \multirow{2}{*}{$\;(aab)$: $C2/m\;$} & \multirow{2}{*}{$\;$\symm{1}{1}; \symm{5}{1}; \symm{7}{1}; \symm{11}{1}}\\
& & & & \\
& & & \multirow{2}{*}{$\;(abc)$: $P\bar{1}\;$} & \multirow{2}{*}{$\;$\symm{1}{1}; \symm{7}{1}}\\
& & & & \\
& & & & \\
& & & & \\
%Gamma5
\hline
& $\sin\mathbf{G}_{M_3}\cdot\mathbf{r}$ & \multirow{2}{*}{${\boldsymbol\epsilon}_5\cos\mathbf{q}_{M_3}\cdot\mathbf{r}$} & \multirow{2}{*}{$\;(aaa)$: $P321\;$} & \multirow{2}{*}{$\;$\symm{1}{1}; \symm{2}{1}; \symm{3}{1};}\\
& $\cos\mathbf{q}_{M_3}\cdot\mathbf{r}$ & & & \\
& \multirow{6}{*}{\raisebox{-0.5\totalheight}{\includegraphics[width=0.25\columnwidth]{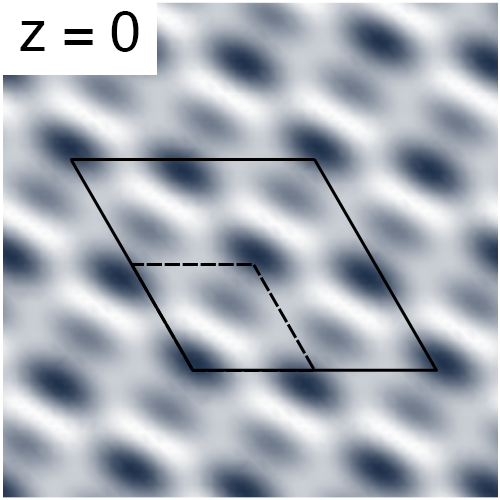}}} & \multirow{6}{*}{\raisebox{-0.5\totalheight}{\includegraphics[width=0.25\columnwidth]{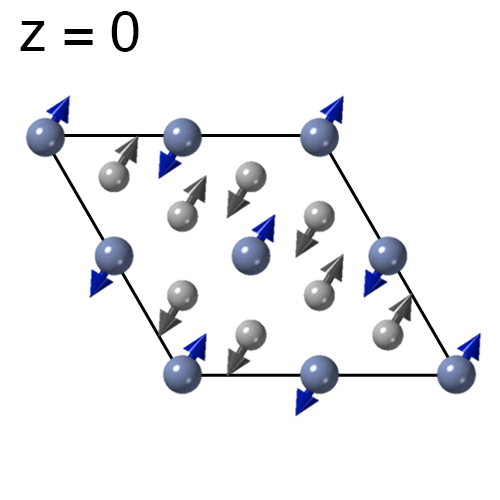}}} & & \multirow{2}{*}{$\;$\symm{4}{1}; \symm{5}{1}; \symm{6}{1};} \\
& & & & \\
& & & & \multirow{2}{*}{$\;$\symm{1}{5}; \symm{2}{5}; \symm{3}{5};} \\
& & & & \\
\multirow{1}{*}{$\;\Gamma_5^-\;$} & & & & \multirow{2}{*}{$\;$\symm{4}{5}; \symm{5}{5}; \symm{6}{5}} \\
\multirow{2}{*}{$(\mathbf{d}_{M}^T)$} & & & & \\
& \multirow{6}{*}{\raisebox{-0.5\totalheight}{\includegraphics[width=0.25\columnwidth]{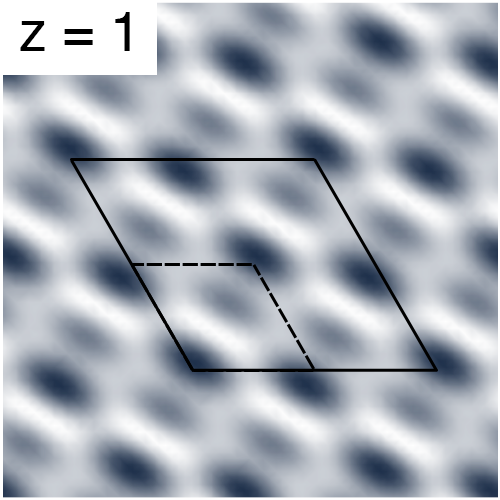}}} & \multirow{6}{*}{\raisebox{-0.5\totalheight}{\includegraphics[width=0.25\columnwidth]{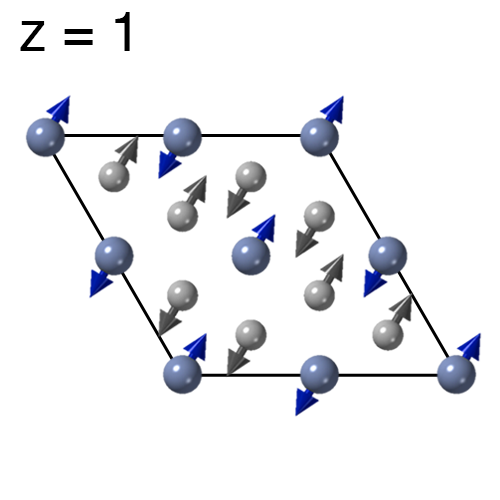}}} & \multirow{2}{*}{$\;(aab)$: $C2\;$} & \multirow{2}{*}{$\;$\symm{1}{1}; \symm{5}{1}; \symm{1}{5}; \symm{5}{5}} \\
& & & & \\
& & & \multirow{2}{*}{$\;(abc)$: $P1\;$} & \multirow{2}{*}{$\;$\symm{1}{1}; \symm{1}{5}} \\
& & & & \\
& & & & \\
& & & & \\
%Gamma7
\hline
& $\sin\mathbf{G}_{L_3}\cdot\mathbf{r}$ & \multirow{2}{*}{${\boldsymbol\epsilon}_7\cos\mathbf{q}_{L_3}\cdot\mathbf{r}$} & \multirow{2}{*}{$\;(aaa)$: $P\bar{3}c1\;$} & \multirow{2}{*}{$\;$\symm{1}{1}; \symm{2}{1}; \symm{3}{1};} \\
& $\cos\mathbf{q}_{L_3}\cdot\mathbf{r}$ & & & \\
& \multirow{6}{*}{\raisebox{-0.5\totalheight}{\includegraphics[width=0.25\columnwidth]{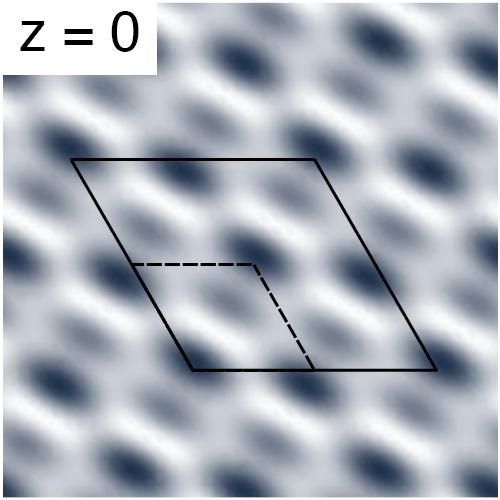}}} & \multirow{6}{*}{\raisebox{-0.5\totalheight}{\includegraphics[width=0.25\columnwidth]{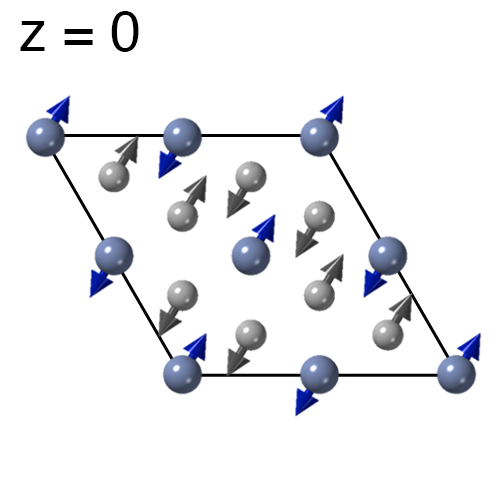}}} & & \multirow{2}{*}{$\;$\symm{4}{1}; \symm{5}{1}; \symm{6}{1};} \\
& & & & \\
& & & & \multirow{2}{*}{$\;$\symm{7}{5}; \symm{8}{5}; \symm{9}{5};} \\
& & & & \\
\multirow{1}{*}{$\;\Gamma_7^-\;$} & & & & \multirow{2}{*}{$\;$\symm{10}{5}; \symm{11}{5}; \symm{12}{5}} \\
\multirow{2}{*}{$(\mathbf{d}_{L}^T)$} & & & & \\
& \multirow{6}{*}{\raisebox{-0.5\totalheight}{\includegraphics[width=0.25\columnwidth]{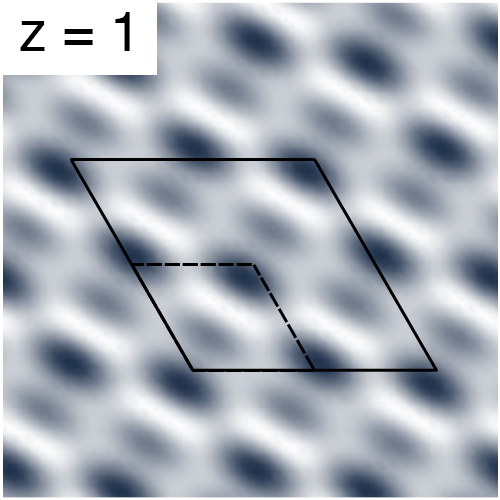}}} & \multirow{6}{*}{\raisebox{-0.5\totalheight}{\includegraphics[width=0.25\columnwidth]{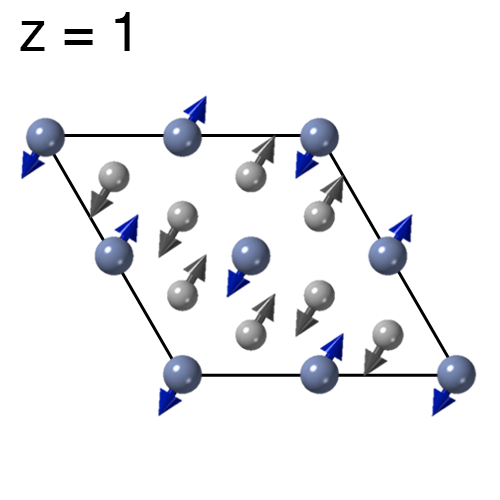}}} & \multirow{2}{*}{$\;(aab)$: $C2/c\;$} & \multirow{2}{*}{$\;$\symm{1}{1}; \symm{5}{1}; \symm{7}{5}; \symm{11}{5}}\\
& & & & \\
& & & \multirow{2}{*}{$\;(abc)$: $P\bar{1}\;$} & \multirow{2}{*}{$\;$\symm{1}{1}; \symm{7}{5}} \\
& & & & \\
& & & & \\
& & & & \\
\hline
\multicolumn{5}{l}{\multirow{2}{*}{{ * $\mathbf{G}_{M_i}$\,=\,$2(\mathbf{q}_{M_{i+1}}-\mathbf{q}_{M_{i+2}})$,\,$\mathbf{G}_{L_i}$\,=\,$2(\mathbf{q}_{L_{i+1}}-\mathbf{q}_{L_{i+2}})$, where $i=1,2,3$ and $i+3=i$.}}} \\
\end{tabular}
%\end{ruledtabular}
\caption{\textbf{The basis functions for selected IRs and their isotropy subgroups.}
The basis functions are plotted for $i$=3 only. For charge, only the lowest-order harmonic, most relevant to valence electrons, is shown for each IR. In our setting, the origin is at the Ti site; note that the inversion center is shifted from the origin by $\mathbf{c}/2$ in the $P\bar{3}c1$ space group, as can be seen from the fact that isotropy subgroups of $\Gamma_7^-$ contains $\{\mathcal{I}|\mathcal{T}_{001} \}$ but not $\{\mathcal{I}|\mathcal{T}_{000} \}$.
}
\label{tab:1}
\end{table*}

\clearpage

\noindent\textbf{Methods}

\noindent\textbf{1$T$-TiSe$_2$ crystals}
1$T$-TiSe$_2$ single crystals grown by chemical vapor transport technique were purchased from HQ graphene. All samples used in this study were prepared from the same single crystal.

\noindent\textbf{Inelastic x-ray scattering}
High-resolution inelastic x-ray scattering experiment was carried out at the 30-IR HERIX beamline of the Advanced Photon Source. The total energy resolution was about 1.5 meV at the incident energy of 23.71 keV using Si (12, 12, 12) reflection of the monochromator. A sample of 50 $\mu$m thickness was mounted on a closed-cycle cryostat and measured in a transmission geometry. Spectra were obtained along high symmetry lines $L-A-\Gamma-M$, which, respectively correspond to $\mathbf{q}$=(2, 1/2, 1/2), (2, 0, 1/2), (2, 0, 0), and (2, 1/2, 0). Lorentzian function with Bose factor was used for fitting:

\begin{multline}
I(\omega) = \frac{A}{1-e^{-\omega/k_{B}T}} \times \\ (-\frac{\Gamma/2}{(\omega+\omega_0)^{2}+(\Gamma/2)^{2}}+\frac{\Gamma/2}{(\omega-\omega_0)^{2}+(\Gamma/2)^{2}}),
\end{multline}

\noindent where $\Gamma$ is the full width at half maximum (FWHM). $\Gamma$ and $\omega_0$ are shown as black markers in Fig.~2e and Fig.~2f, respectively.

\noindent\textbf{Raman spectroscopy}
Raman spectroscopy was performed on a home-built instrument equipped with 750-mm spectrometer and liquid-nitrogen-cooled CCD detector using 633-nm laser as an excitation source. Bragg grating notch filters were used to suppress elastic signals, allowing measuring inelastic signals above 5 cm$^{-1}$ with a resolution of 0.33 cm$^{-1}$. Parallel ($\parallel$) and crossed ($\perp$) polarizations of the incident and scattered lights propagating along the $c$-axis were measured in a back-scattering geometry. The incident light polarization was rotated about the $c$-axis using half-wave plates. The laser power and beam-spot size are kept below 0.5 mW and 2 $\mu$m, respectively, which gave laser heating of 10 K estimated based on the Stokes and anti-Stokes intensity ratio.

\noindent\textbf{X-ray diffraction}
Prior to inelastic x-ray scattering measurements, samples were tested for consistency with earlier works using x-ray diffraction (XRD). A XRD series on a representative 1$T$-TiSe$_2$ sample prepared from the same single-crystal was collected on a STOE imaging plate diffraction system (IPDS-2T) using Mo $K_\alpha$ radiation. For the investigated specimen all accessible reflections ($\approx21000$) were measured up to a maximum angle of $2\Theta=65^\circ$. The data were corrected for Lorentz, polarization, extinction, and absorption effects. Using SHELX \cite{she08} and JANA2006 \cite{pet14}, all averaged symmetry-independent reflections ($I>2\sigma$) have been included for the refinements. For all temperatures the unit cell and the space group (SG) were determined, the atoms were localized in the unit cell utilizing random phases as well as Patterson superposition methods, the structure was completed and solved using difference Fourier analysis, and finally the structure was refined. In all cases the refinements converged quite well and showed excellent reliability factors (see GOF, $R_1$, and $wR_2$ in Supplementary Table 5).

Additional XRD experiments were performed at the 1C beamline of the Pohang Light Source using reflection geometry with an incident energy of 11.216 keV to search for weak forbidden reflections. To reduce the background, a Si (8, 4, 4) bent analyzer was positioned between the sample and the detector, which gave a signal-to-noise ratio better than $10^7$ \cite{kim23}.

\noindent\textbf{Data availability}

All data are available in the main text or the supplementary information.

\noindent\textbf{Acknowledgements}

We thank Jasper van Wezel and Alaska Subedi for insightful discussions. This project is supported by IBS-R014-A2 and National Research Foundation (NRF) of Korea through the SRC (No.~2018R1A5A6075964). The use of the Advanced Photon Source at the Argonne National Laboratory was supported by the U. S. DOE under Contract No. DE-AC02-06CH11357. C.K. acknowledges the support by the Institute for Basic Science in Korea (Grant No. IBS-R009-G2, IBS-R009-D1) and the National Research Foundation (NRF) of Korea grant funded by the Korea government (MSIT) (No. 2022R1A3B1077234). J.M.B. acknowledges the support by the National Research Foundation (NRF) of Korea (No. 2022R1C1C2008671). K.-S.K. is supported by the Ministry of Education, Science, and Technology (No. 2021R1A2C1006453, 2021R1A4A3029839) of the National Research Foundation (NRF) of Korea.

\noindent\textbf{Author contributions}

B.J.K. conceived and managed the project. H.-W.J.K., K.K. and A.S. performed inelastic x-ray scattering experiments. K.K. and Ho.K. performed Raman scattering experiments. S.H., K.K., J.-K.K., Ja.K., Hy.K. and Ju.K. performed synchrotron-based x-ray diffraction experiments. Ho.K., J.S., S.J. and C.K. performed second harmonic generation experiments. M.M. and M.L.T performed structure refinement using x-ray diffraction. K.K., H.-W.J.K., S.H. and B.J.K. performed representation analysis. J.M.B. and K.-S.K. performed RPA calculations. K.K., H.-W.J.K., S.H., J.M.B., K.-S.K. and B.J.K. wrote the manuscript with inputs from all authors.

\noindent\textbf{Competing interests}

Authors declare that they have no competing interests.

\end{document}